\documentclass[aps,prl,twocolumn,showpacs,superscriptaddress]{revtex4}

\usepackage[dvips]{color}     
\usepackage{graphicx}

\begin{document}

\title{Potential energy surfaces for cluster emitting nuclei}

\author{Dorin N. Poenaru}
\email[]{poenaru@th.physik.uni-frankfurt.de}
%\homepage[]{http://idranap.nipne.ro/Poenaru/Poenaru.htm}
\affiliation{
Horia Hulubei National Institute of Physics and Nuclear
Engineering, \\RO-077125 Bucharest-Magurele, Romania}
\affiliation{Frankfurt Institute for Advanced Studies,
J. W. Goethe Universit\"at, 
        Max-von-Laue-Str. 1,    D-60438 Frankfurt am Main,   Germany}

\author{Radu A. Gherghescu}
\affiliation{
Horia Hulubei National Institute of Physics and Nuclear
Engineering, \\RO-077125 Bucharest-Magurele, Romania}
\affiliation{Frankfurt Institute for Advanced Studies,
J. W. Goethe Universit\"at, 
        Max-von-Laue-Str. 1,    D-60438 Frankfurt am Main,   Germany}

\author{Walter Greiner}
\affiliation{Frankfurt Institute for Advanced Studies,
J. W. Goethe Universit\"at, 
        Max-von-Laue-Str. 1,    D-60438 Frankfurt am Main,   Germany}

\date{\today}

\begin{abstract}
Potential energy surfaces are calculated by using the
most advanced asymmetric two-center
shell model allowing to obtain shell and pairing corrections which are added
to the Yukawa-plus-exponential model deformation energy. Shell effects are
of crucial importance for 
experimental observation of spontaneous disintegration by heavy ion
emission. Results for $^{222}$Ra, $^{232}$U, $^{236}$Pu and
$^{242}$Cm illustrate the main ideas
and show for the first time for a cluster emitter a potential barrier 
obtained by using the macroscopic-microscopic method.
\end{abstract}


\maketitle

\section{Introduction}
Recently we performed a systematic analysis \cite{p240pr02} of the
experimental results concerning heavy particle radioactivities
\cite{p195b96b} showing that our predictions within the analytical
superasymmetric (ASAF) model (see \cite{p160adnd91} and the references therein) 
have been confirmed and that the strong shell
effects of the daughter $^{208}$Pb were not fully exploited. In this way we
could make suggestions for the candidates to be used in the future experiments.
%\vspace*{-170mm}
%\noindent
%Invited talk at the Carpathian summer school on {\em Exotic Nuclei and
%Nuclear/Particle Astrophysics}, Mamaia, June 13-24, 2005.  
%\vspace{136mm}

In the present work we take advantage of using the most advanced two center
shell model \cite{rad03prc} to study the potential energy surfaces (PES) of
cluster emitting nuclei showing deep valleys due to the doubly magic
heavy fragments $^{208}$Pb and $^{132}$Sn. The Strutinsky's \cite{str67np} 
macroscopic-microscopic method is used. A particularly deep valley
is that of $^{208}$Pb which proved to be of practical importance not only for
the production of superheavy nuclei but also for experimental search of
cluster decay modes. Even for alpha decay it is possible to see such a
valley if the emitter is $^{212}$Po or $^{106}$Te. In the later case 
the heavy fragment
$^{102}$Sn plays the important role. The potential barrier shape for a heavy ion
decay mode may be obtained by cutting the PES at a given value of the mass
and charge asymmetry. In this way one can compare the difference between the
macroscopic barrier and the total one with shell and pairing corrections
taken into account,  providing a further justification of the ASAF
barrier shape.

\section{Macroscopic energy}
In a binary fission process $^A Z \rightarrow \ ^{A_1}Z_1 \ + \ ^{A_2}Z_2$ the 
phenomenological energy $E_{Y+EM}$ is calculated within Yukawa-plus-exponential 
model (Y+EM) \cite{sch69zp,kra79pr} by taking into account
the difference between charge and mass asymmetry \cite{p80cpc80}. 

By requesting zero deformation energy for a spherical shape, the
potential energy is defined as
\begin{eqnarray}
E_{Y+EM} & = (E_Y - E_Y^0)+(E_c - E_c^0) \nonumber \\  & = E_Y^0[B_Y - 1 +
2X(B_c -1)]
\end{eqnarray}
where 
$E_Y^0=a_2A^{2/3} \{ 1-3x^2+(1+1/x)[2+3x(1+x)]\exp (-2/x) \}
$,
 $E_c^0 = a_cZ^2A^{-1/3}$ are 
energies corresponding to spherical shape and  $a_2=a_s(1-\kappa I^2)$,
$I=(N-Z)/A$, $x=a/R_0$, $R_0=r_0A^{1/3}$.
The parameters $a_s , \kappa , a_c=3e^2/(5r_0)$, and $r_0$ are taken 
from M\" oller et al.~\cite{mol95adndt}.

The relative Yukawa and Coulomb energies
$B_Y=E_Y/E_Y^0$, $B_c=E_c/E_c^0$ are functions of the nuclear
shape. The dependence on the neutron and proton numbers is contained
in $E_Y^0$, in the fissility parameter $X=E_c^0/(2E_Y^0)$ and $B_Y$.
For  a binary fragmentation with charge 
densities $\rho_{1e}$ and $\rho_{2e}$, one has \cite{p80cpc80} a relative
energy
\begin{equation}
B_Y =\frac{E_Y}{E_Y ^0} = \frac{a_{21}}{a_{20}} B_{Y1} +
\frac{\sqrt {a_{21} a_{22}}}{a_{20}} B_{Y12} +
\frac{a_{22}}{a_{20}} B_{Y2}
\end{equation}
with axially-symmetric shape-dependent terms expressed by triple integrals
\begin{equation}
B_{Y1}= b_Y \int _{-1}^{x_c} dx \int _{-1} ^{x_c} dx'\int_0^1dwF_1F_2Q_Y 
\end{equation}
\begin{equation}
B_{Y12}= b_Y\int _{-1} ^{x_c} dx \int _{x_c} ^{1} dx'\int_0^1dwF_1F_2Q_Y 
\end{equation}
\begin{equation}
B_{Y2}= b_Y \int _{x_c} ^{1} dx \int _{x_c} ^{1} dx'\int_0^1dwF_1F_2Q_Y 
\end{equation}
in which $b_Y=-d^4(r_0/2a^2)a_2R_0A/E_Y^0$, $d = (z'' - z')/2R_0$ is the
nuclear semilength in units of $R_0$ and
\begin{equation} 
F_1=y^2+yy_1 \cos \varphi - \frac{x-x'}{2}\frac{dy^2}{dx}
\end{equation}
\begin{eqnarray} 
Q_Y= & \{[\sqrt{P}(\sqrt{P}+2a/R_0d)+2a^2/(R_0d)^2]\cdot \nonumber \\  & 
\exp (-R_0\sqrt{P}d/a)-2a^2/(R_0d)^2\}/P^2
\end{eqnarray} 
$F_2$ is obtained from $F_1$ by replacing $dy^2/dx$ with
$dy_1^2/dx'$. In the above equations $P=y^2+y_1^2-2yy_1 \cos \varphi+
(x-x')^2$, $w=\varphi /2\pi$,   and $x_c$ is the
position of separation plane between fragments with -1, +1 intercepts
on the symmetry axis (surface equation $y = y(x)$ or $y_1 = y(x')$).
The integrals are computed
numerically by Gauss-Legendre quadratures.

In a similar way the Coulomb relative energy is given by
\begin{equation}
B_c =\frac{E_c}{E_c ^0} = \left ( \frac{\rho _{1e}}{\rho _{0e}}
\right ) ^2 B_{c1} + \frac{\rho _{1e} \rho _{2e}}{\rho _{0e} ^2}
B_{c12} + \left ( \frac{\rho _{2e}}{\rho _{0e}} \right ) ^2 B_{c2} 
\end{equation}
and for axially symmetric shapes
\begin{equation}
B_{c1}= b_c \int _{-1} ^{x_c} dx \int _{-1} ^{x_c} dx' F(x, x') 
\end{equation}
\begin{equation}
B_{c12}= b_c \int _{-1} ^{x_c} dx \int _{x_c} ^{1} dx' F(x, x') 
\end{equation}
\begin{equation}
B_{c2}= b_c \int _{x_c} ^{1} dx \int _{x_c} ^{1} dx' F(x, x') 
\end{equation}
where $b_c = 5d^5 /8\pi$.
In the integrand
\begin{eqnarray} 
F(x,x') = & \{ y y_1[(K-2D)/3]\cdot  \nonumber \\ &
\left [ 2(y^2+y_1^2)-(x-x')^2+ \right . \nonumber \\ & \left .
\frac{3}{2}(x
-x')\left ( \frac{dy_1^2}{dx'}-\frac{dy^2}{dx} \right ) \right ] +
\nonumber \\ &
K \left \{ y^2y_1^2/3+\left [y^2-\frac{x-x'}{2}\frac{dy^2}{dx}
\right ]\cdot \right . \nonumber \\ & \left .
\left [y_1^2-\frac{x-x'}{2}\frac{dy_1^2}{dx'}\right ] 
\right \} \} a_{\rho}^{-1} 
\end{eqnarray}  
$K$ and $K'$ are the complete elliptic integrals of the first and
second kind, respectively:
\begin{equation}
K(k) = \int _0^{\pi /2}(1-k^2 {\sin}^2 t)^{-1/2} dt 
\end{equation}
\begin{equation}
K'(k) = \int _0^{\pi /2}(1-k^2 {\sin}^2 t)^{1/2} dt 
\end{equation}
and $a_{\rho} ^2 = (y+y_1)^2+(x-x')^2$, $k^2 = 4yy_1 /a_{\rho}^2$, $D
= (K - K')/k^2$. 
The elliptic integrals may be calculated by using Chebyshev polynomial
approximation. For $x = x'$ the function $F$ is not determined. In
this case, after removing the indetermination, we get $F(x,x')=4y^3
/3$.

Starting from the touching point configuration, $R \geq R_t$,
for spherical shapes of the
fragments,  one can use {\em analytical relationships.}
The Coulomb interaction energy of a system of two spherical nuclei,
separated by a distance $R$ between centers, is $E_{c12} = e^2 Z_1
Z_2 /R$, where $e$ is the electron charge. 

Within a liquid drop model (LDM) there is no 
contribution of the surface energy to
the interaction of the separated fragments; the barrier has 
a maximum at the touching point configuration.
The proximity forces acting at small separation distances (within the
range of strong interactions) give rise in the Y+EM to an interaction
term expressed as folllows
\begin{eqnarray}
E_{Y12} =& -4\left ( \frac{a}{r_0} \right ) ^2 \sqrt {a_{21} a_{22}}
\frac{\exp (- R/a)}{R/a} \nonumber \\  & \cdot
\left [ g_1 g_2 \left ( 4+\frac{R}{a} \right ) -g_2f_1 - g_1f_2
\right ]
\end{eqnarray}
where
\begin{equation}
g_k =  \frac{R_k}{a} \cosh \left ( \frac{R_k}{a}  \right ) - \sinh
\left ( \frac{R_k}{a} \right ) 
\end{equation}
\begin{equation}
f_k = \left (\frac{R_k}{a}  
\right ) ^2 \sinh \left ( \frac{R_k}{a} \right )
\end{equation}
In many cases the interaction energy is maximum at a certain distance
$R_m > R_t = R_1 + R_2$, which can be found by solving numerically
the following nonlinear equation
\begin{equation}
e^x + p_1 + x(p_1 + xp) = 0 \; ; \; x=R/a
\end{equation}
in which
\begin{equation}
p = - \frac{a^3}{r_0^2} \sqrt {a_{21} a_{22}} \frac{g_1 g_2}{e^2 Z_1
Z_2} 
\end{equation}
\begin{equation}
p_1 = p(4-f_1 /g_1 -f_2 /g_2 )
\end{equation}
and the interval $x_t =R_t/a, x_t +5$ may be given as input data of a 
program using M\"uller's iteration scheme of succesive bisections and 
inverse parabolic interpolation.

\section{Shell and pairing corrections}

In the following we would like to outline the calculations of the shell
\cite{str67np} and pairing \cite{bol72pr} corrections 
$\delta E = \delta U + \delta P$ leading to the total deformation energy
\begin{equation}
E_{def}(R,\eta)=E_{Y+EM}(R,\eta) + \delta E(R,\eta)
\end{equation} 
More details are given in the book \cite{p195b96b}.
By choosing two intersected spheres for nuclear shape parametrization
one can take the separation distance between fragment centers, $R$, as a 
deformation parameter. Initially, for a parent nucleus $R_i=R_0 - R_2$. 
At the touching point $R_t=R_1 + R_2$. The mass asymmetry
$\eta=(A_1-A_2)/A$.

The two-center shell model \cite{rad03prc} gives at every pair of
coordinates $(R,\eta)$ the sequence of doubly
degenerate discrete energy levels $\epsilon _i = E_i /\hbar \omega^0_0$ in
units of $\hbar \omega^0_0 = 41 A^{-1/3}$, arranged in order of increasing
energy. The smoothed-level distribution density is
obtained by averaging the actual distribution over a finite energy
interval $\Gamma = \gamma  \hbar \omega _0^0$, with $\gamma \simeq 1$,
\begin{eqnarray}
\tilde{g}(\epsilon) = & \left \{ \sum_{i=1}^{n_m} [2.1875+y_i(y_i(1.75-y_i/6)
\right . \nonumber \\  & \left .
-4.375)]e^{-y_i} \right \}(1.77245385\gamma)^{-1} 
\end{eqnarray}
where $y=x^2=\left [(\epsilon  - \epsilon _i)/\gamma \right ]^2$.
The summation is performed up to the level $n_m$ fulfilling the
condition $|x_i|\geq 3$.

The Fermi energy, $\tilde{\lambda}$, of this distribution is given by
\begin{equation}
N_p=2\int_{-\infty}^{\tilde{\lambda}} \tilde{g}(\epsilon) 
d \epsilon
\end{equation}
with $N_p=Z$ for proton levels and $N_p=A-Z$ for neutron levels, leading to
a non-linear equation in $\tilde{\lambda}$, solved numerically. 
The total energy of the uniform level distribution
\begin{equation}
\tilde{u} = \tilde{U}/\hbar \omega_0^0 = 
2\int_{-\infty}^{\tilde{
\lambda}} \tilde{g}(\epsilon)\epsilon d \epsilon
\end{equation}
In units of $\hbar \omega_0^0$ the shell corrections are calculated
for each pair $(R,\eta)$: 
\begin{equation}
\delta u(n, R,\eta) = \sum_{i=1}^n 2\epsilon_i(R,\eta) -
\tilde{u}(n, R,\eta)
\end{equation}
$n=N_p/2$ particles. Then $\delta u =
\delta u_p + \delta u_n$. 
\begin{figure}[ht]
\includegraphics[width=8cm]{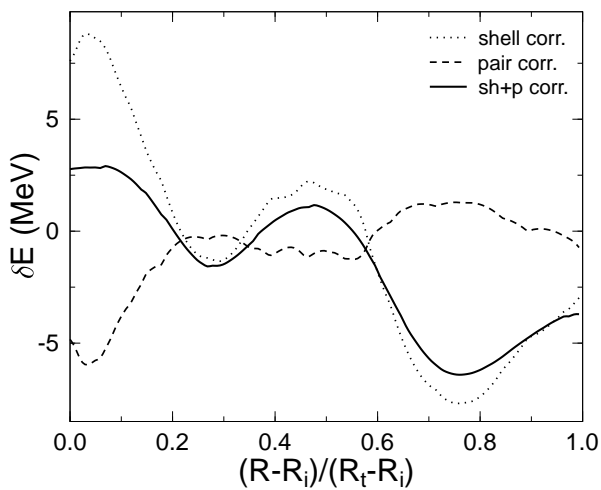}
\caption{Shell and pairing corrections for a symmetric ($\eta=0$)
fission of $^{236}$Pu.  
\label{pu}}
\end{figure}

Similarly, for pairing corrections we take the doubly
degenerate levels $\{\epsilon_i\}$ in units of $\hbar \omega_0^0$.
$Z/2$ levels are occupied with $n$ levels below and $n'$ above
Fermi energy contributing to pairing, $n=n'=\Omega \tilde{g_s} /2$.
The cutoff energy, $\Omega \simeq 1 \gg \tilde{\Delta}=12/\sqrt{A}\hbar\omega_0^0$.
The gap $\Delta$ and Fermi energy $\lambda$ are solutions of the BCS system
of two  eqs:
\begin{equation}
0 = \sum_{k_i}^{k_f}\frac{\epsilon_k -\lambda}{\sqrt{(\epsilon_k 
-\lambda)^2+\Delta^2}}
\end{equation}
\begin{equation}
 \frac{2}{G} =
\sum_{k_i}^{k_f}\frac{1}{\sqrt{(\epsilon
_k -\lambda)^2+\Delta^2}}
\end{equation}
where $k_i=Z/2-n+1, \; \; k_f=Z/2+n'$, and
\begin{equation}
\frac{2}{G} \simeq 2\tilde{g}(\tilde{\lambda})\ln \left (\frac
{2\Omega}{\tilde{\Delta}} \right )
\end{equation}

As a consequence of the pairing correlation, the levels
below the Fermi energy are only partially filled, while those above
the Fermi energy are partially empty.
Occupation probability by a quasiparticle ($u_k$) or hole ($v_k$) is given
by
\begin{equation}
v_k^2 = \left
[1-(\epsilon_k -\lambda)/E_k \right ]/2 ; \; \; u_k^2 = 1-v_k^2
\end{equation}
The quasiparticle energy is expressed as
\begin{equation}
E_\nu=\sqrt{(\epsilon _\nu -\lambda)^2+\Delta^2}. 
\end{equation}
The pairing correction $\delta p=p-\tilde{p}$,
represents the difference between the pairing correlation
energies for the discrete level distribution
\begin{equation}
p = \sum_{k=k_i}^{k_f}2v_k^2\epsilon_k -2\sum_{k=k_i}^{Z/2}\epsilon_k
- \frac{\Delta^2}{G}
\end{equation}
and for the continuous level distribution
\begin{equation}
\tilde{p} = -(\tilde{g}\tilde{\Delta^2})/2 =-(\tilde{g}_s\tilde{\Delta
^2})/4
\end{equation}
Compared to shell correction, the pairing correction is out of phase
and smaller (see Fig.\ref{pu}) leading for $\eta=$~constant
to a smoother total curve $\delta e (R) =
\delta u (R) + \delta p (R) $ where $\delta p = \delta p_p + \delta p_n$.
\begin{figure}[ht]
\begin{center}
\includegraphics[width=8cm]{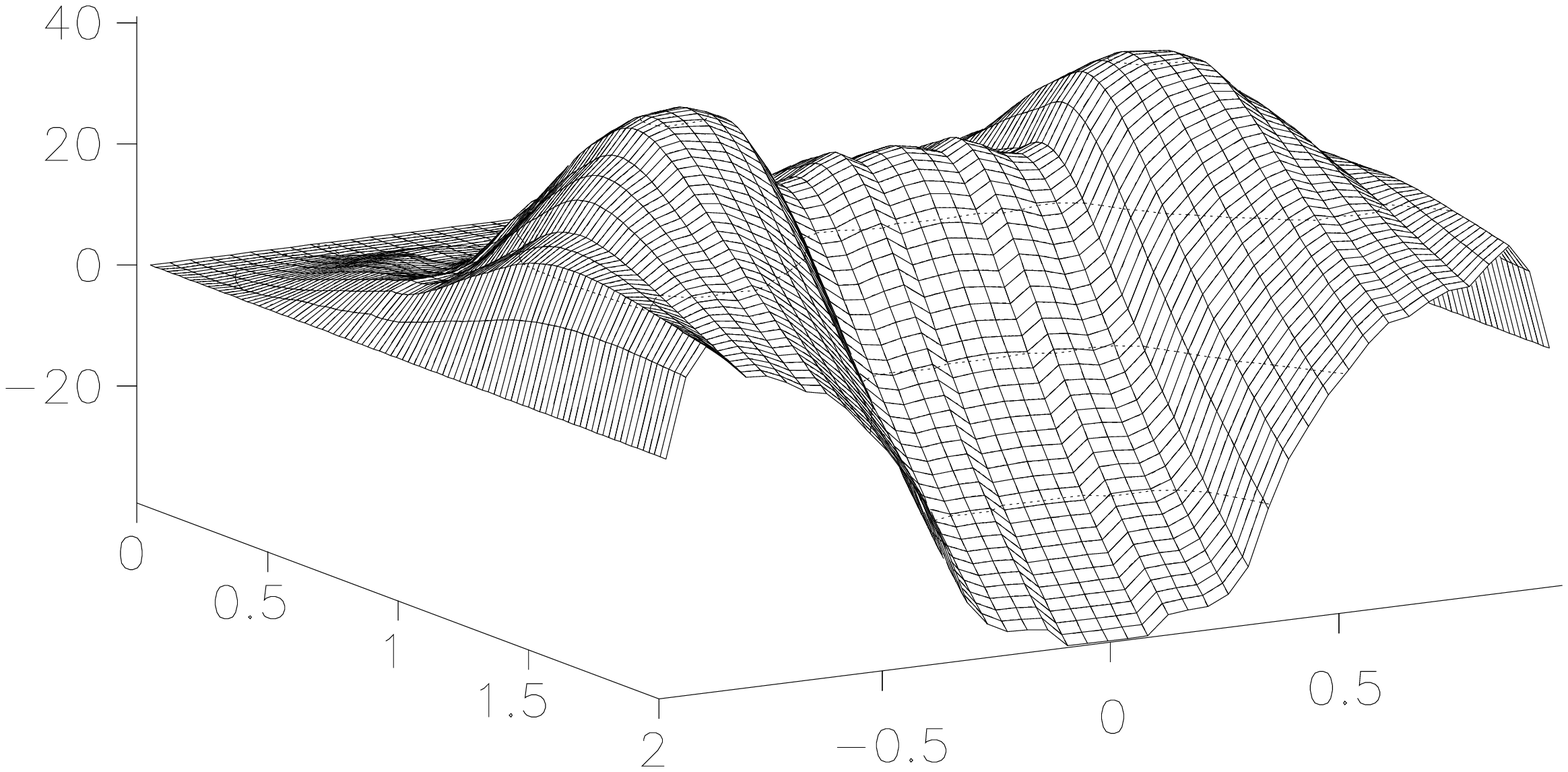} \vspace*{-0.9cm}

\includegraphics[width=8cm]{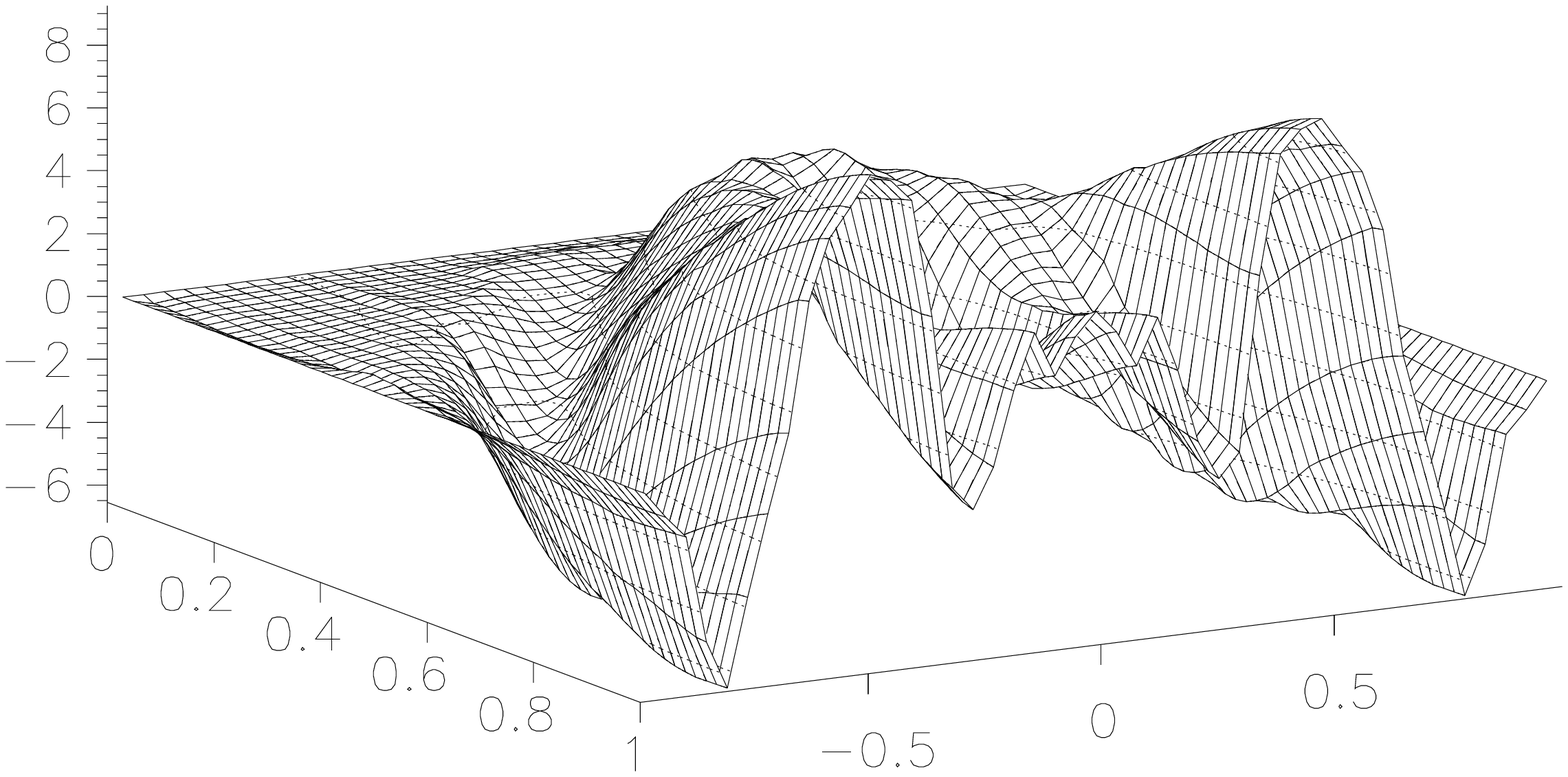} \vspace*{-0.9cm}

\includegraphics[width=8cm]{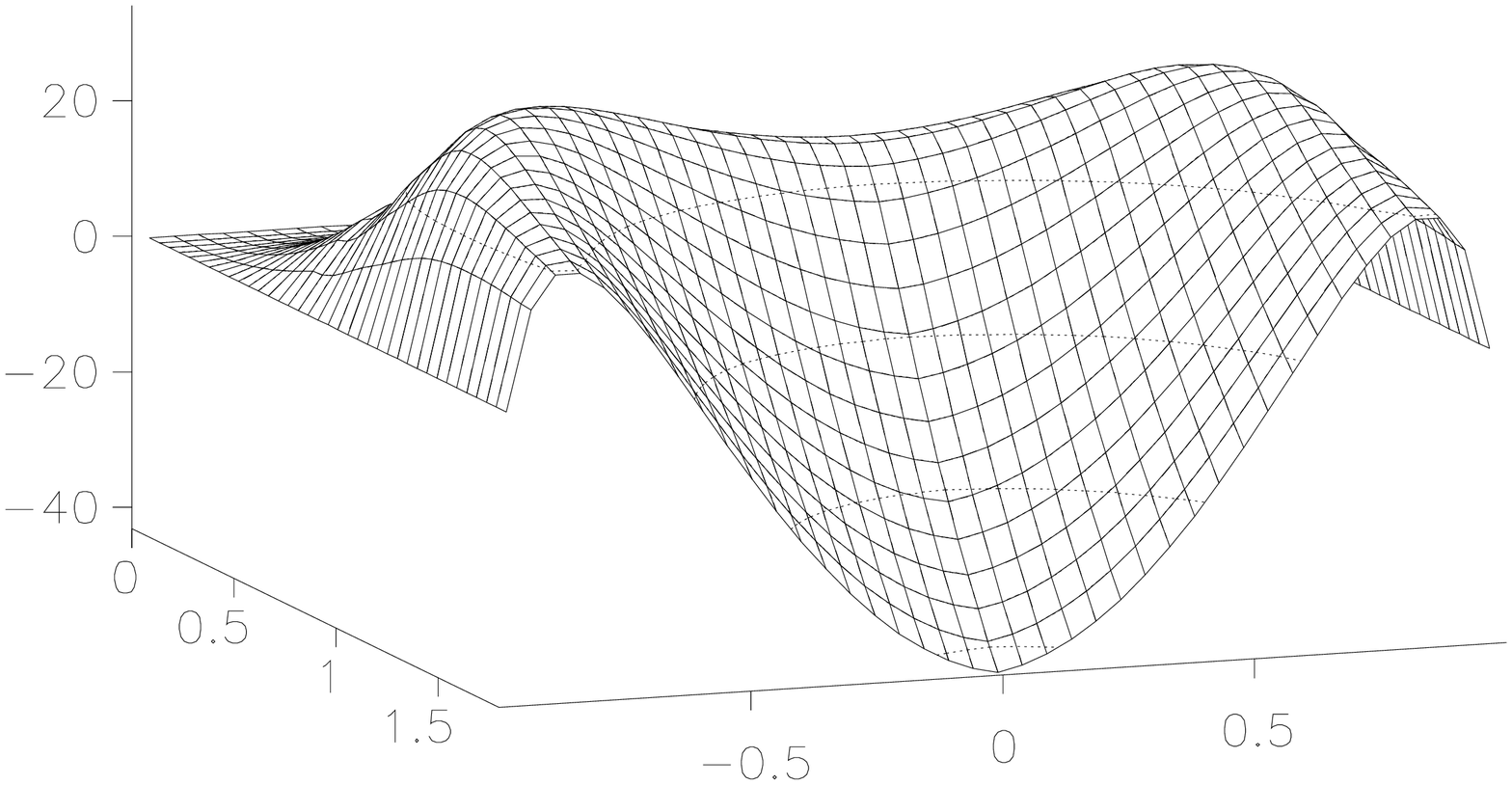}  
\end{center}
\caption{PES of $^{222}$Ra vs $(R-R_i)/(R_t - R_i) \geq 0$ and $\eta =
(A_1-A_2)/(A_1+A_2)$. Y+EM (bottom), Shell + Pairing corrections (center), and
total deformation energy (top). The energies are 
expressed in MeV. \label{pesra}}
\end{figure}

\section{Results}
In the following we shall present results for $^{222}$Ra, $^{232}$U,
$^{236}$Pu, and $^{242}$Cm, which are emitters of $^{14}$C, $^{24}$Ne,
$^{28}$Mg, and $^{34}$Si, respectively. All have been experimentally 
observed. In order to obtain a relatively smooth PES we made the
approximation $\eta=\eta_A=(A_1-A_2)/A \simeq \eta_Z=(Z_1-Z_2)/Z$. In this
way the fragment nucleon numbers $N_1, Z_1$ and $N_2, Z_2$ (plotted in
figures \ref{ranzt}, \ref{unzt}, \ref{punzt}, \ref{cmnzt}) are linear
functions of $\eta$.
We prefer to use the dimensionless separation distance
$\xi=(R-R_i)/(R_t-R_i)$ instead of $R$. In this way one can clearly see the
initial parent nucleus at $\xi=0$ and the touching point configuration at
$\xi=1$.
We adopt the usual convention of having zero
deformation energy and shell plus pairing corrections for the initial
spherical shape, leading to $E_{def} = E_{Y+EM} = \delta E =0$ at $R=R_i$
for all values of
$\eta$ and at $\eta =\pm 1$ for all values of $R$.

\subsection{$^{222}$Ra}
The PES versus the normalized separation
distance $(R-R_i)/(R_t - R_i)$ and the mass asymmetry $\eta =
(A_1-A_2)/(A_1+A_2)$ are plotted in Fig. \ref{pesra}. The macroscopic Y+EM
deformation energy is shown at the bottom, followed by the microscopic shell 
plus pairing corrections (center), and their sum (the total deformation
energy) at the top. 

\begin{figure}[ht]
\includegraphics[width=6cm]{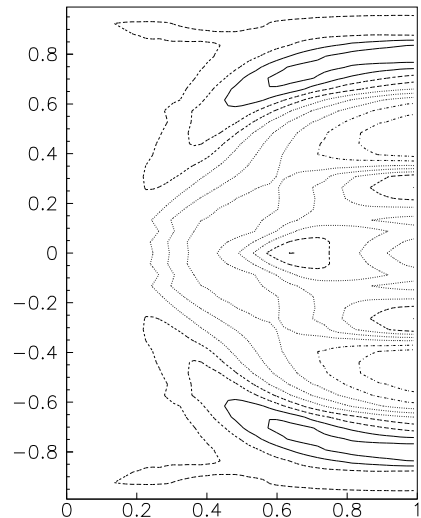}      
\caption{Contour plot of shell and pairing corrections for $^{222}$Ra  vs
$(R-R_i)/(R_t - R_i)$ and $\eta$.
\label{contra}}
\end{figure}

\begin{figure}[hbt]
\includegraphics[width=6cm]{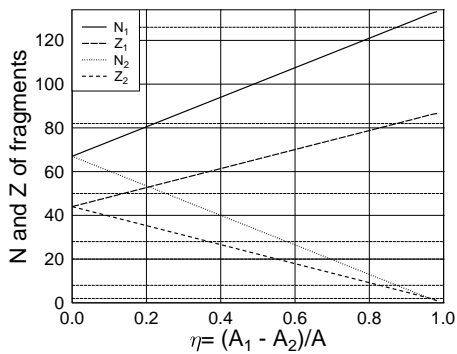}
\caption{Variation of the neutron and proton numbers of the two 
fragments with the mass asymmetry at the touching point, $R=R_t$ for
$^{222}$Ra. \label{ranzt}}
\end{figure}

Three valleys around $\eta \simeq 0.8$; $0.3$ and $0.1$
can be seen in the center of Fig. \ref{pesra} and
on the corresponding contour plot (Fig.~\ref{contra}). 
We only count the number of valleys for $\eta \geq 0$ because
the mirror $\eta \leq 0$ gives the same number for complimentary heavy
fragments becoming light ones and vice-versa.
In figure~\ref{ranzt}
we shall see that they are produced due to the magicity of the nucleon
number of the fragments.
Such cold valleys were used in
the sixtieth by Walter Greiner to motivate the search for superheavy
elements, and the development of Heavy Ion Physics worldwide and in Germany,
where GSI was built. These valleys may be also seen on the total PES at the
top of Fig.~\ref{pesra}. Here the deepest valley remains that at a small
value of $\eta$ not far from the minimum of the macroscopic Y+EM energy at
$\eta=0$, which is responsible for  the
\begin{figure}[htb]
\includegraphics[width=6cm]{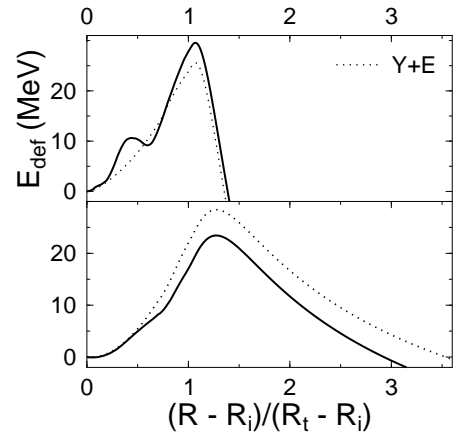}
\caption{A cut through the PES of $^{222}$Ra at symmetry $\eta =
0$ (top) and for $^{14}$C radioactivity with $^{208}$Pb daughter (bottom). 
\label{barra}}
\end{figure}
\begin{figure}[htb]
\includegraphics[width=6cm]{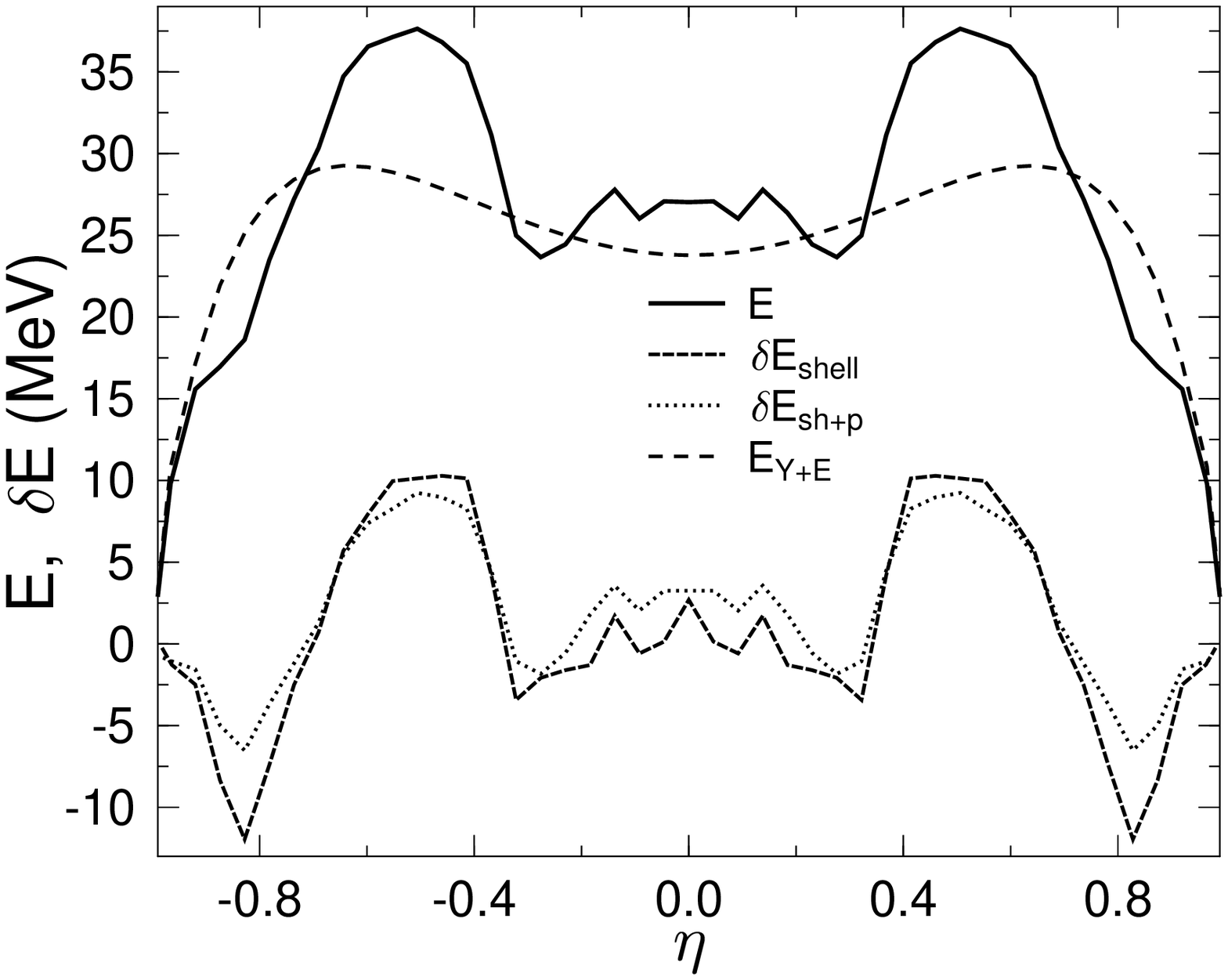}
\caption{Deformation energies at the touching point configurations ($R=R_t$)
of $^{222}$Ra vs the asymmetry $\eta $: $E_{Y+EM}$ and $E_{def}$ (top); shell
+ pairing corrections and only shell corrections (bottom). \label{ratouc}}
\end{figure}
cold fission. At a large value of $\eta$, 
the $^{208}$Pb + $^{14}$C
valley, favouring the $^{14}$C radioactivity of $^{222}$Ra, is laying 
on the Businaro-Gallone mountain, hence it is shallower despite the fact
that it is very pronounced in the shell correction surface. 
On the contour plot (figure~\ref{contra}) one can see how it evolves from
lower mass asymmetry at a small value of separation distance, $\xi$, to the
larger $\eta$ at the touching point, $\xi=1$. 

Two plots obtained by cutting the total 
PES and that of Y+EM at a given value of the asymmetry parameter are
shown in Fig.~\ref{barra}. 
In the upper part one can see a two humped barrier
at $\eta =0$. The typical example for $^{14}$C emission from $^{222}$Ra,
shown at the bottom, provide
justification for one of the basic assumption of the analytical
superasymmetric fission model, which was very successful in
predicting the half-lives of cluster decay modes.
It is remarkable to see for the first time for a cluster emitter a potential
barrier obtained by using the macroscopic-microscopic method. Having a
smaller height and width compared to the (dotted line) macroscopic Y+EM
barrier, it is very similar to the barrier used in ASAF which was lower and
narrower than the Myers-Swiatecki's \cite{mye66np} LDM barrier.

One can see how deep are the two main valleys on the PES by plotting in
Fig.~\ref{ratouc} a cut of the PES at the touching point configuration,
$R=R_t$. In the upper part of this figure the macroscopic energy,
$E_{Y+E}$ (the smooth dashed line), and the total deformation energy, $E$
(heavy line) can be seen. 
In the lower part of Fig.~\ref{ratouc} there are two other curves
representing the shell plus pairing correction (dotted curve) and only the
shell corrections (dashed line). As we already mentioned when the
Fig.~\ref{pu} was discussed, there are two effects of pairing 
corrections leading to a smoother variation and to a shallower valley.
\begin{figure}[ht]
\begin{center}
\includegraphics[width=8cm]{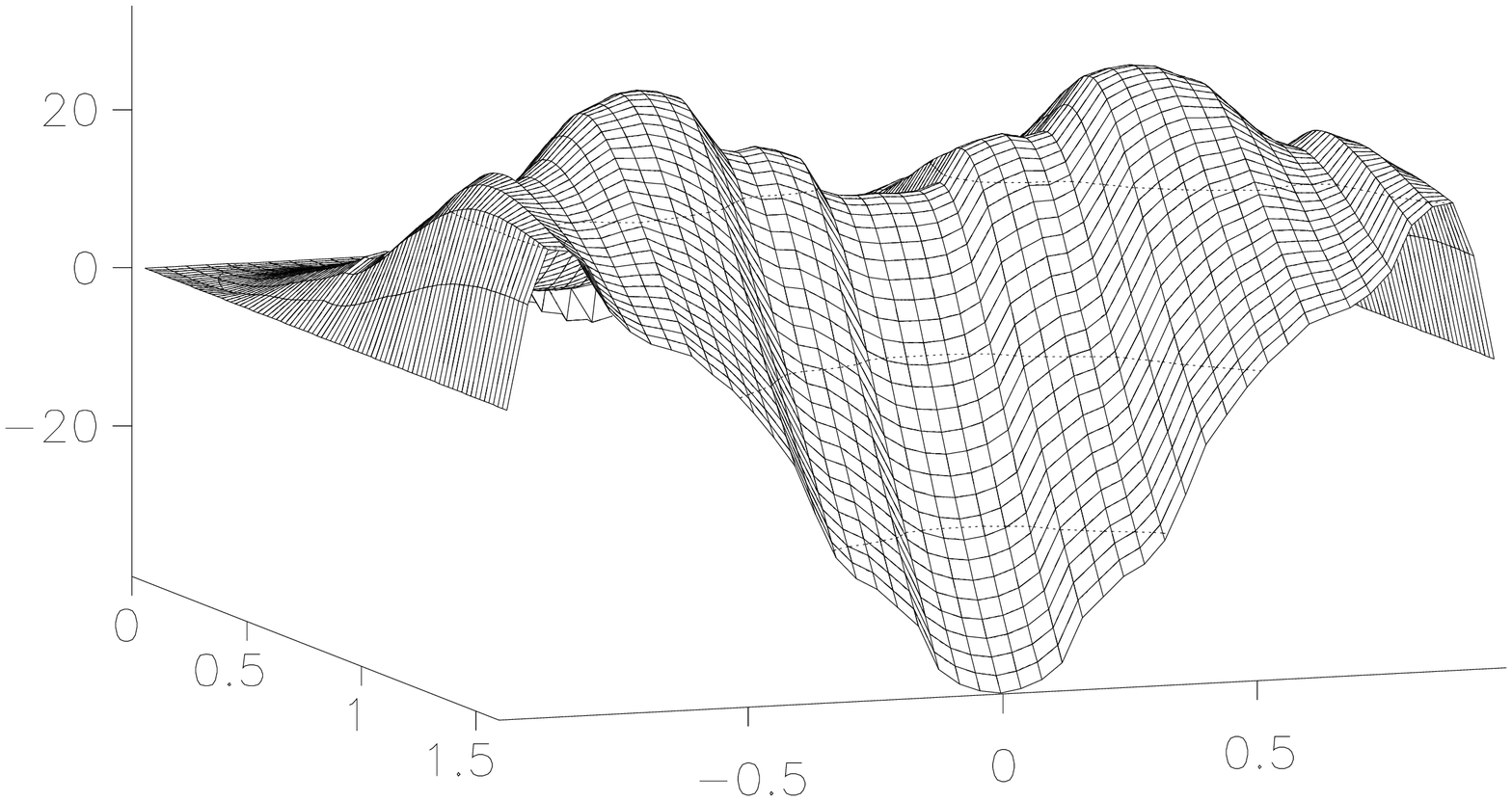} \vspace*{-0.9cm}

\includegraphics[width=8cm]{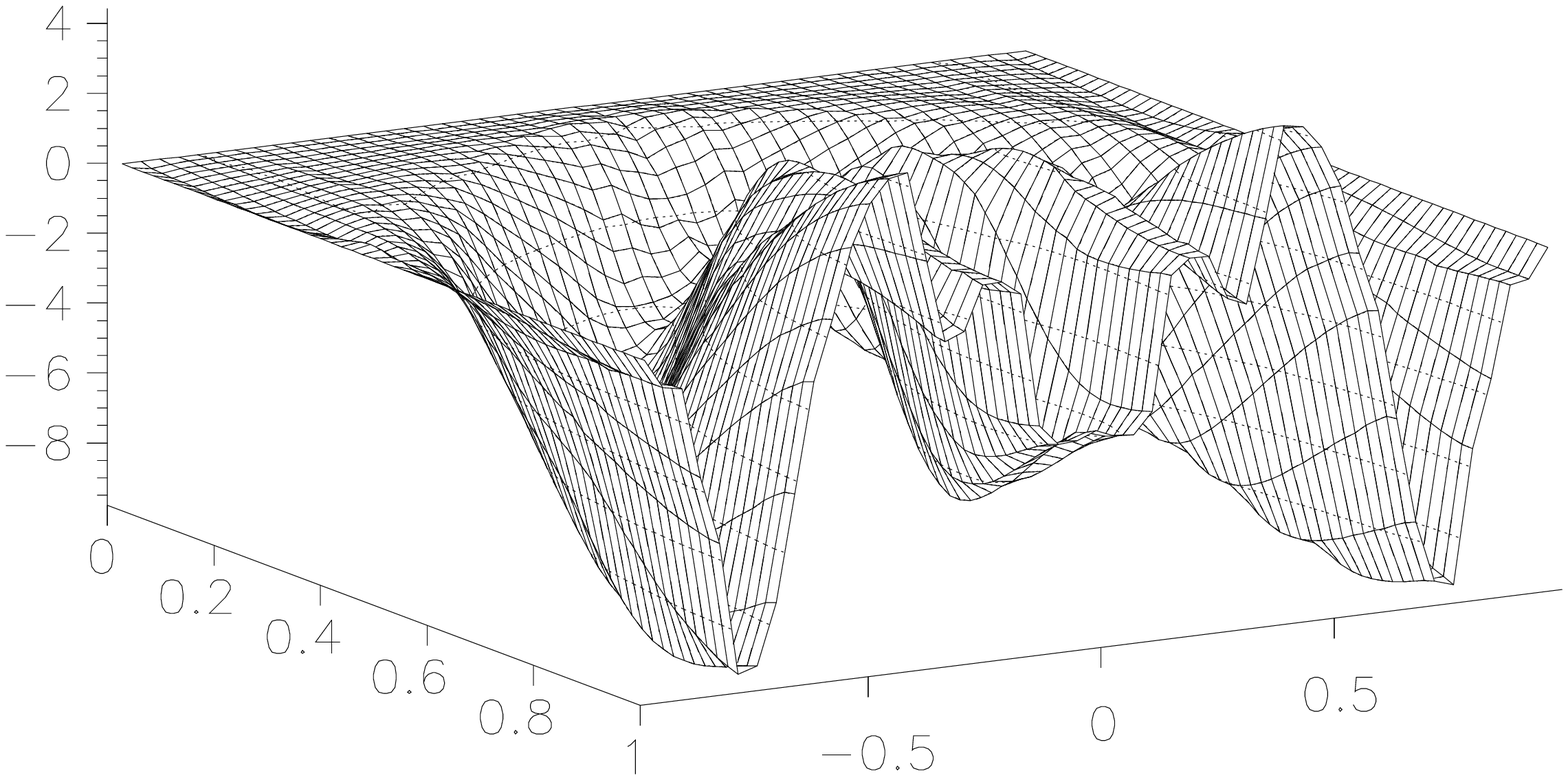} \vspace*{-0.9cm}

\includegraphics[width=8cm]{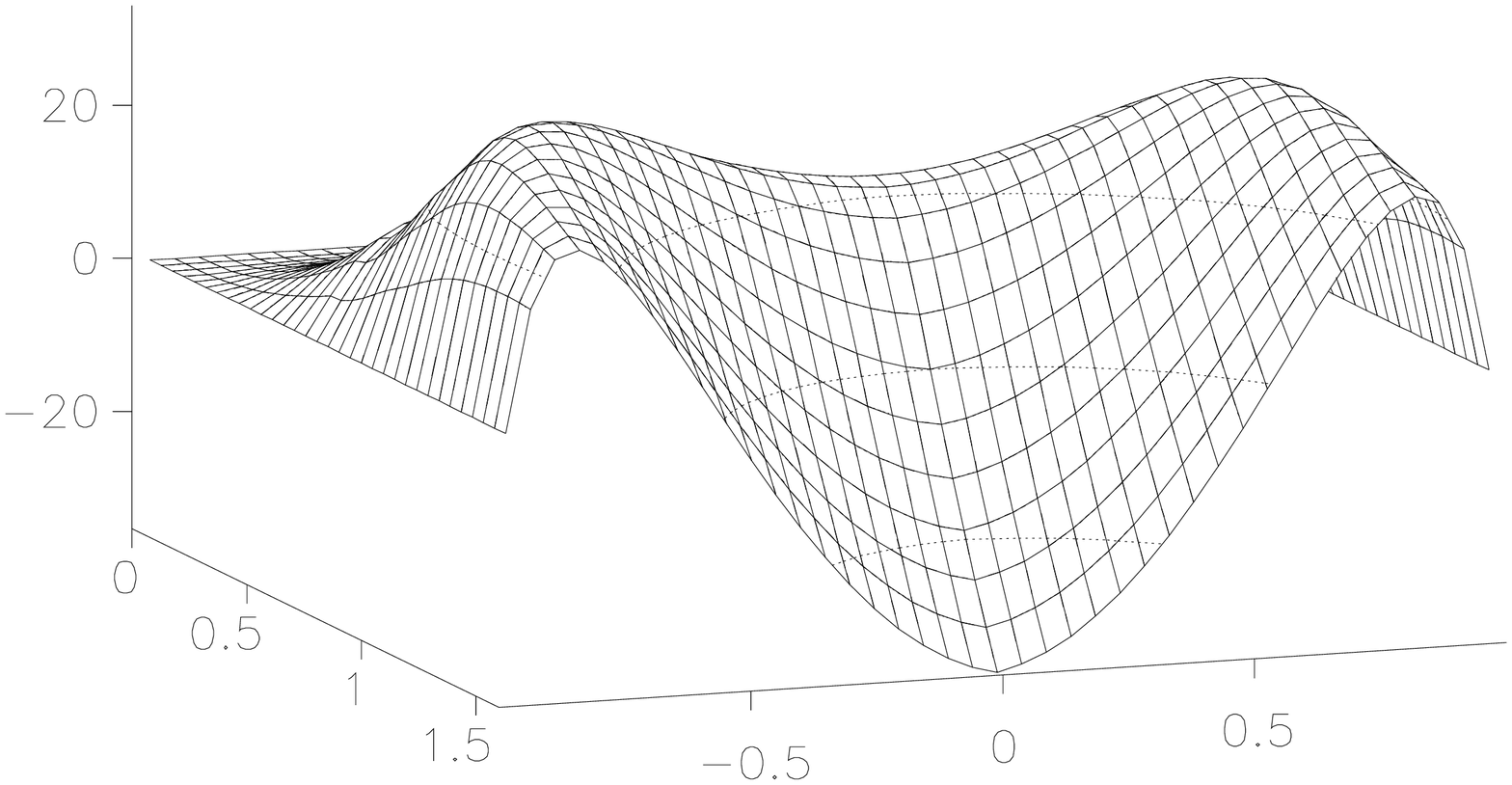}  
\end{center}
\caption{PES of $^{232}$U vs $(R-R_i)/(R_t - R_i)$ and $\eta =
(A_1-A_2)/(A_1+A_2)$. Y+EM (bottom), Shell + Pairing corrections (center), and
total deformation energy (top). The energies are 
expressed in MeV. \label{pesu}}
\end{figure}

We can better understand how the deep valleys appear in figure~\ref{ratouc}
if we plot in figure~\ref{ranzt} the variation of the neutron and proton
numbers of the two
fragments with the mass asymmetry at the touching point, $R=R_t$, for
$^{222}$Ra. Every time a nucleon number reaches a magic value, the
corresponding shell correction has a local minimum. The very deep valley
at $\eta > 0.8$ (for different mass and charge asymmetry at
$\eta_A=0.874$ and $\eta_Z=0.864$)
is produced by reaching almost simultaneously three magic
numbers $N_1=126$,  $N_2=8$, and $Z_1=82$ (the decay $^{222}$Ra $\rightarrow
\ ^{208}$Pb + $^{14}$C). The next one, at an intermediate value of $\eta$
($\eta_A=0.369$, $\eta_Z=0.363$), is mainly due to the Ni~($Z_2=28$) light fragment
($^{222}$Ra $\rightarrow \ ^{152}$Nd + $^{70}_{28}$Ni$_{42}$). Finally the valley at a small
asymmetry parameter ($\eta_A=0.189$ and $\eta_Z=0.136$) corresponds to the cold
fission process ($^{222}$Ra $\rightarrow \ ^{132}$Sn + $^{90}$Sr). 
\begin{figure}[ht]
\includegraphics[width=6cm]{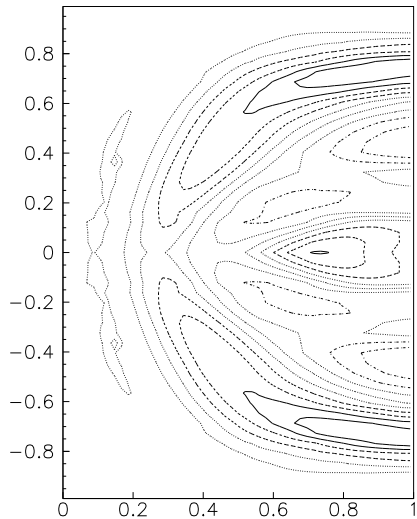}      
\caption{Contour plot of shell and pairing corrections for $^{232}$U  vs
$(R-R_i)/(R_t - R_i)$ and $\eta$.
\label{contu}}
\end{figure}

\begin{figure}[htb]
\includegraphics[width=6cm]{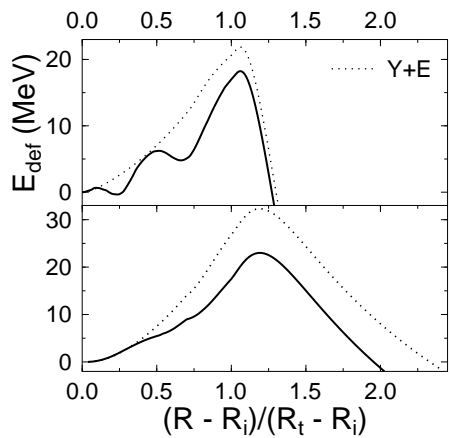}
\caption{A cut through the PES of $^{232}$U at symmetry $\eta =
0$ (top) and for $^{24}$Ne radioactivity with $^{208}$Pb daughter (bottom). 
\label{baru}}
\end{figure}

\subsection{$^{232}$U}
Three PES versus the normalized separation distance 
and the mass asymmetry 
are plotted in Fig. \ref{pesu}: the macroscopic Y+EM
deformation energy at the bottom; the microscopic shell 
plus pairing corrections at the center, and their sum (the total deformation
energy) at the top. There are again three valleys on the shell plus pairing
corrections and on the contour plot (Fig.~\ref{contu}) 
around $\eta \simeq 0.8$; $0.3$, and $0.15$. Unlike on the figure \ref{pesra}
for $^{222}$Ra in which the valley due to Sn at a low mass asymmetry is not
so deep, now this cold fission valley is well shaped and the trend continues
for heavier nuclei such as $^{236}$Pu and $^{242}$Cm. An intermediate valley
is produced by the magic neutron number $N_2=50$ of the light fragment.
They are also present on the total PES at the
top of Fig.~\ref{pesu}. Here the deepest valley remains that of Sn,
which is responsible for the cold fission; the $^{208}$Pb + $^{24}$Ne
valley, explaining the $^{24}$Ne radioactivity of $^{232}$U, is laying 
on the Businaro-Gallone mountain, hence it is shallower.
\begin{figure}[h]
\includegraphics[width=6cm]{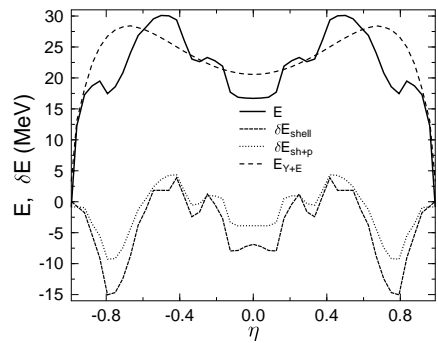}
\caption{Deformation energies at the touching point configurations ($R=R_t$)
of $^{232}$U vs the asymmetry $\eta $: $E_{Y+EM}$ and $E_{def}$ (top); shell
+ pairing corrections and only shell corrections (bottom). \label{utouc}}
\end{figure}
\begin{figure}[h]
\includegraphics[width=6cm]{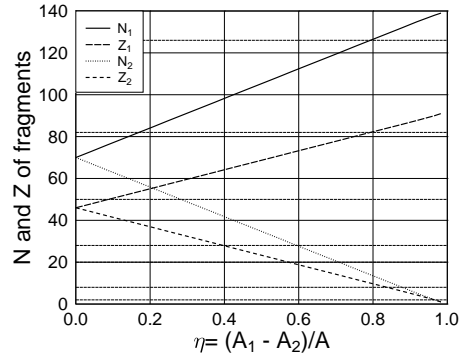}
\caption{Variation of the neutron and proton numbers of the two 
fragments with the mass asymmetry at the touching point, $R=R_t$ for
$^{232}$U. \label{unzt}}
\end{figure}

In the upper part of Fig.~\ref{baru} one can see a two humped barrier
at $\eta =0$. The example for $^{24}$Ne emission from $^{232}$U,
shown at the bottom, provide
justification for one of the basic assumption of the analytical
superasymmetric fission model already mentioned in the previous
subsection.

A cut of the PES at the touching point configuration,
$R=R_t$ is shown in Fig.~\ref{utouc} with the $E_{Y+EM}$ (the smooth dashed
line) and the total deformation energy $E_{def}$ (heavy line) in
the upper part. The three valleys mentioned above are present both on the
total deformation energy (top) and
in the lower part where the shell plus pairing correction (dotted curve) and 
only the shell corrections (dashed line) are plotted.
\begin{figure}[ht]
\begin{center}
\includegraphics[width=8cm]{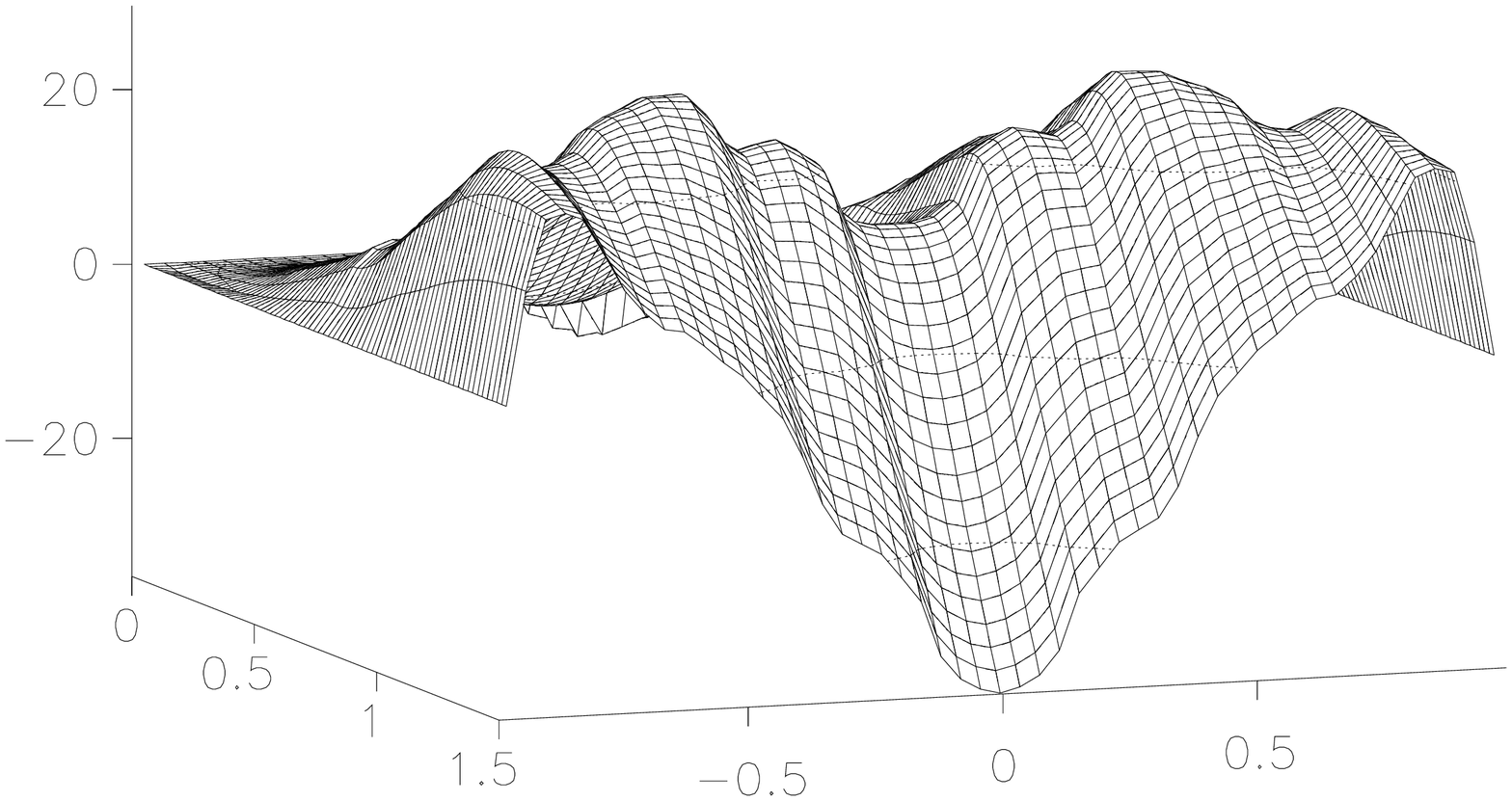} \vspace*{-0.9cm}

\includegraphics[width=8cm]{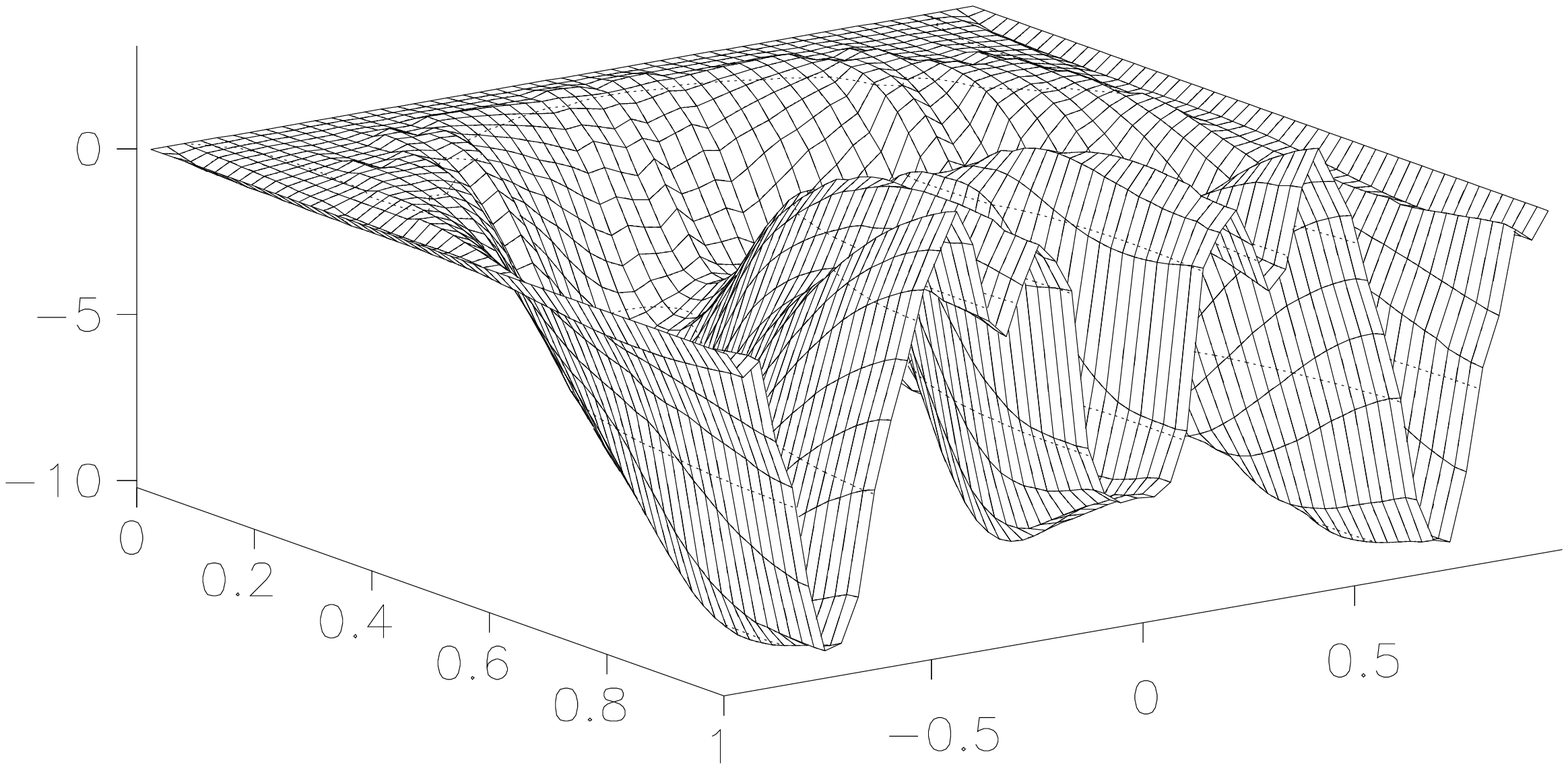} \vspace*{-0.9cm}

\includegraphics[width=8cm]{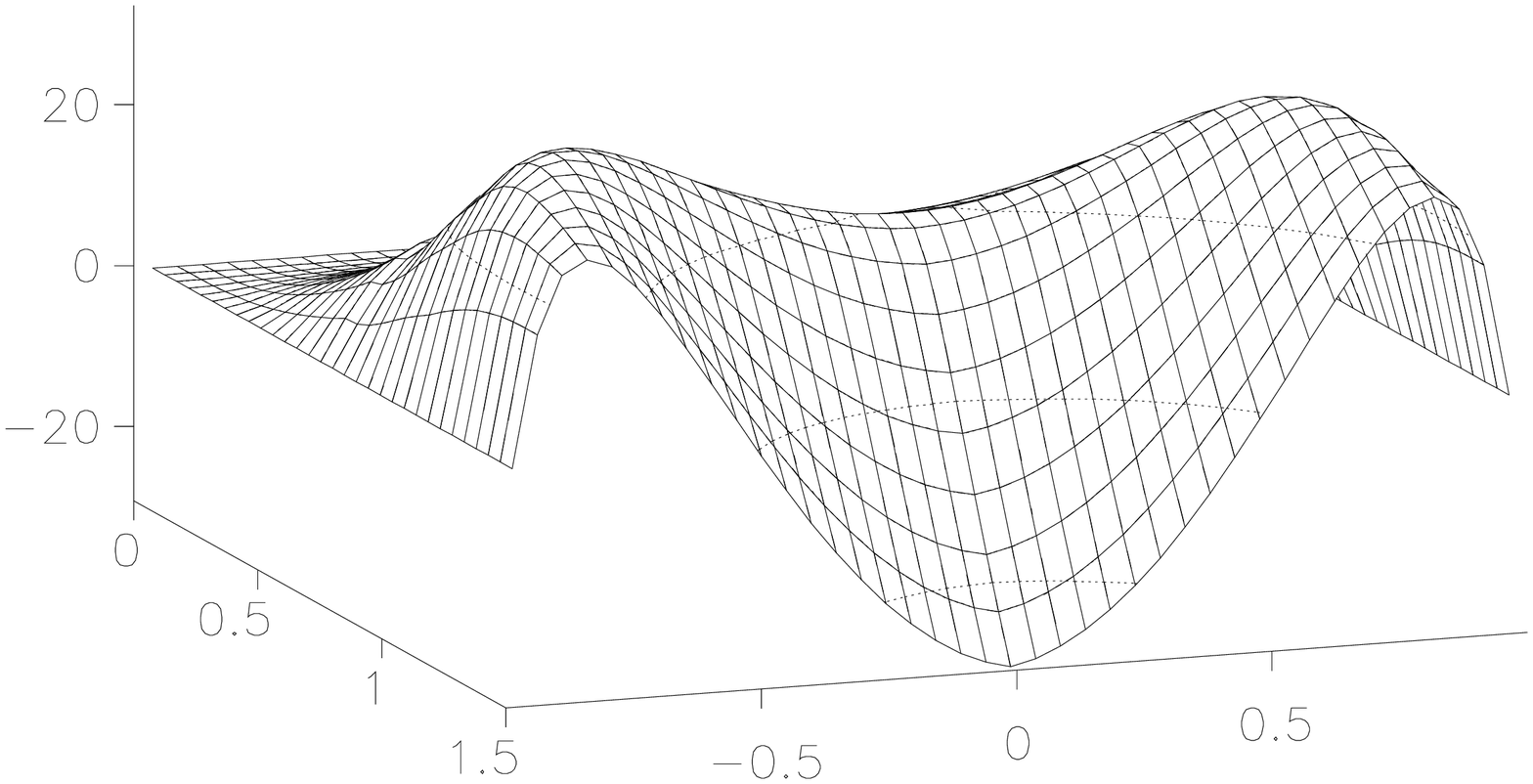}  
\end{center}
\caption{PES of $^{236}$Pu vs $(R-R_i)/(R_t - R_i)$ and $\eta =
(A_1-A_2)/(A_1+A_2)$. Y+EM (bottom), Shell + Pairing corrections (center), and
total deformation energy (top). The energies are 
expressed in MeV. \label{pespu}}
\end{figure}

The figure~\ref{unzt} showing the variation of the neutron and proton
numbers of the two
fragments with the mass asymmetry at the touching point, $R=R_t$, for
$^{232}$U allows us to understand how the deep valleys from figure~\ref{utouc} are
produced. The very deep valley around $\eta =0.8$ ($\eta_A =0.793$, $\eta_Z
=0.783$) is produced by the negative shell corrections 
due to two magic numbers $N_1=126$ and $Z_1=82$ ($^{24}$Ne radioactivity
$^{232}$U $\rightarrow \
^{208}$Pb + $^{24}$Ne). Another doubly magic heavy fragment, $^{132}$Sn,
produces the cold fission valley ($^{232}$U $\rightarrow \ ^{132}$Sn +
$^{100}$Mo) at low mass and charge asymmetry ($\eta_A =0.138$, $\eta_Z
=0.087$). 
\begin{figure}[ht]
\includegraphics[width=6cm]{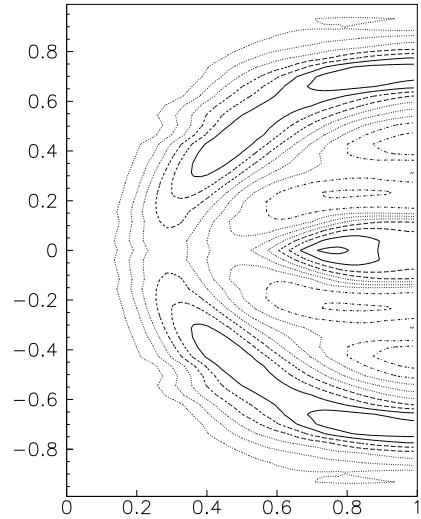}      
\caption{Contour plot of shell and pairing corrections for $^{236}$Pu  vs
$(R-R_i)/(R_t - R_i)$ and $\eta$.
\label{contpu}}
\end{figure}
\begin{figure}[htb]
\includegraphics[width=6cm]{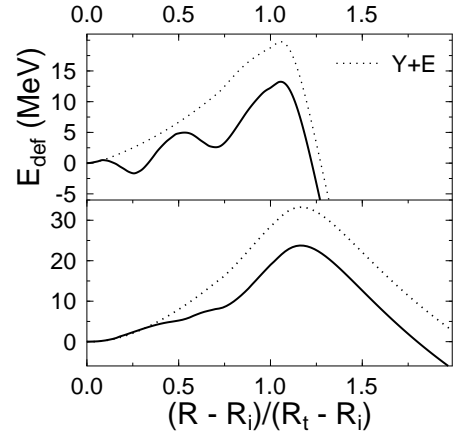}
\caption{A cut through the PES of $^{236}$Pu at symmetry $\eta =
0$ (top) and for $^{28}$Mg radioactivity with $^{208}$Pb daughter (bottom). 
\label{barpu}}
\end{figure}

The intermediate one ($\eta_A =0.293$, $\eta_Z=0.304$), quite shallow,
is due to the neutron magic number of the light fragment $N_2=50$ ($^{232}$U
$\rightarrow \ ^{150}$Nd + $^{82}_{32}$Ge$_{50}$).

\subsection{$^{236}$Pu}

PES versus the normalized separation
distance and the mass asymmetry are plotted in Fig. \ref{pespu}, where
Y+EM deformation energy is shown at the bottom, the shell 
plus pairing corrections at the center, and their sum (the total deformation
energy) at the top. Very deep valleys due to the doubly magic fragments 
$^{208}$Pb and $^{132}$Sn can be seen in the center of Fig. \ref{pespu} and
on the corresponding contour plot (Fig.~\ref{contpu}) around $\eta \simeq 0.75$ 
and $\eta < 0.1$, respectively. At an intermediate value of the asymmetry
parameter, $\eta \simeq 0.28$ there is another valley not so deep.
These valleys may be also seen on the total PES at the
top of Fig.~\ref{pespu}. Here the deepest valley remains that of Sn,
which is responsible for the cold fission; the $^{208}$Pb + $^{28}$Mg
valley, explaining the $^{28}$Mg radioactivity of $^{236}$Pu, is laying 
on the Businaro-Gallone mountain, hence it is shallower.

Two plots obtained by cutting the PES at a given value of the asymmetry are
shown in Fig.~\ref{barpu}: a two humped barrier at $\eta =0$ in the upper part
and the barrier for $^{24}$Ne spontaneous emission from $^{236}$Pu,
shown at the bottom. The last one provides a qualitative microscopic
justification for the chosen barrier shape within the phenomenological ASAF.
\begin{figure}[htb]
\includegraphics[width=6cm]{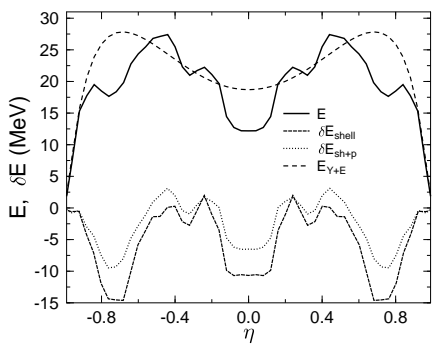}
\caption{Deformation energies at the touching point configurations ($R=R_t$)
of $^{236}$Pu vs the asymmetry $\eta $: $E_{Y+EM}$ and $E_{def}$ (top); shell
+ pairing corrections and only shell corrections (bottom). \label{putouc}}
\end{figure}
\begin{figure}[hbt]
\includegraphics[width=6cm]{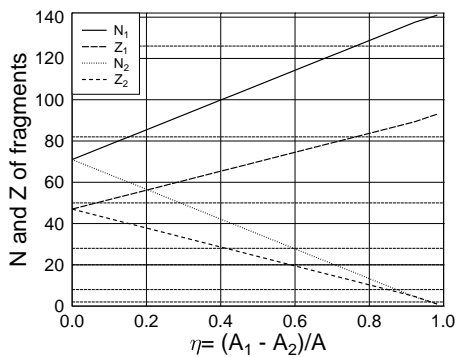}
\caption{Variation of the neutron and proton numbers of the two 
fragments with the mass asymmetry at the touching point, $R=R_t$ for
$^{236}$Pu. \label{punzt}}
\end{figure}

One can see how deep are the three valleys by plotting in
Fig.~\ref{putouc} a cut of the PES at the touching point configuration,
$R=R_t$. In the upper part of this figure the macroscopic energy,
$E_{Y+E}$ (the smooth dashed line), and the total deformation energy, $E_{def}$
(heavy line) can be seen. 
They exhibit few valleys of which that 
of Sn at lower values of
$\eta$ and that of Pb at higher $\eta$ are the deepest ones.
In the lower part of this figure there are two other curves
representing the shell plus pairing correction (dotted curve), $\delta
E_{sh+p}$, and only the shell corrections (dashed line), $\delta E_{shell}$. 
The valley at the intermediate value of the
asymmetry parameter, $\eta \simeq 0.28$, is quite shallow.
\begin{figure}[ht]
\begin{center}
\includegraphics[width=8cm]{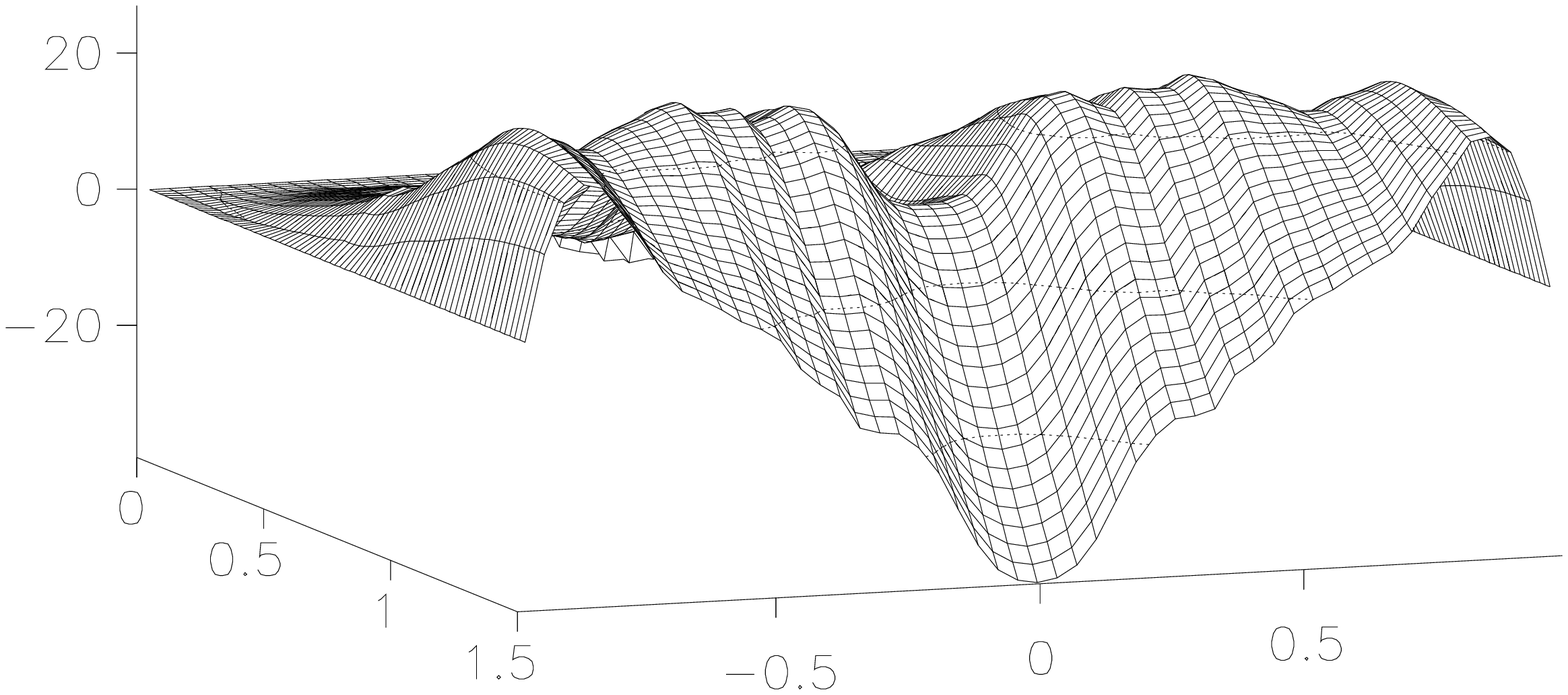} \vspace*{-0.9cm}

\includegraphics[width=8cm]{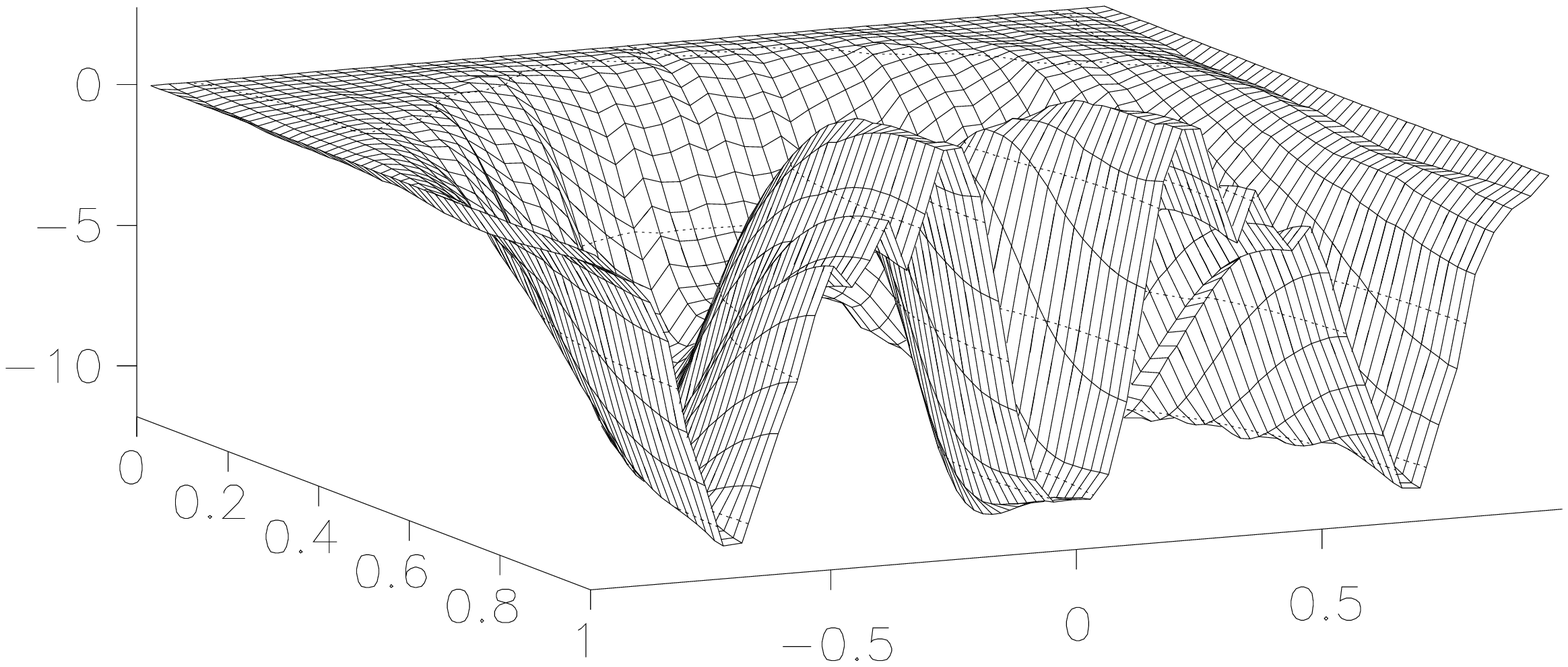} \vspace*{-0.9cm}

\includegraphics[width=8cm]{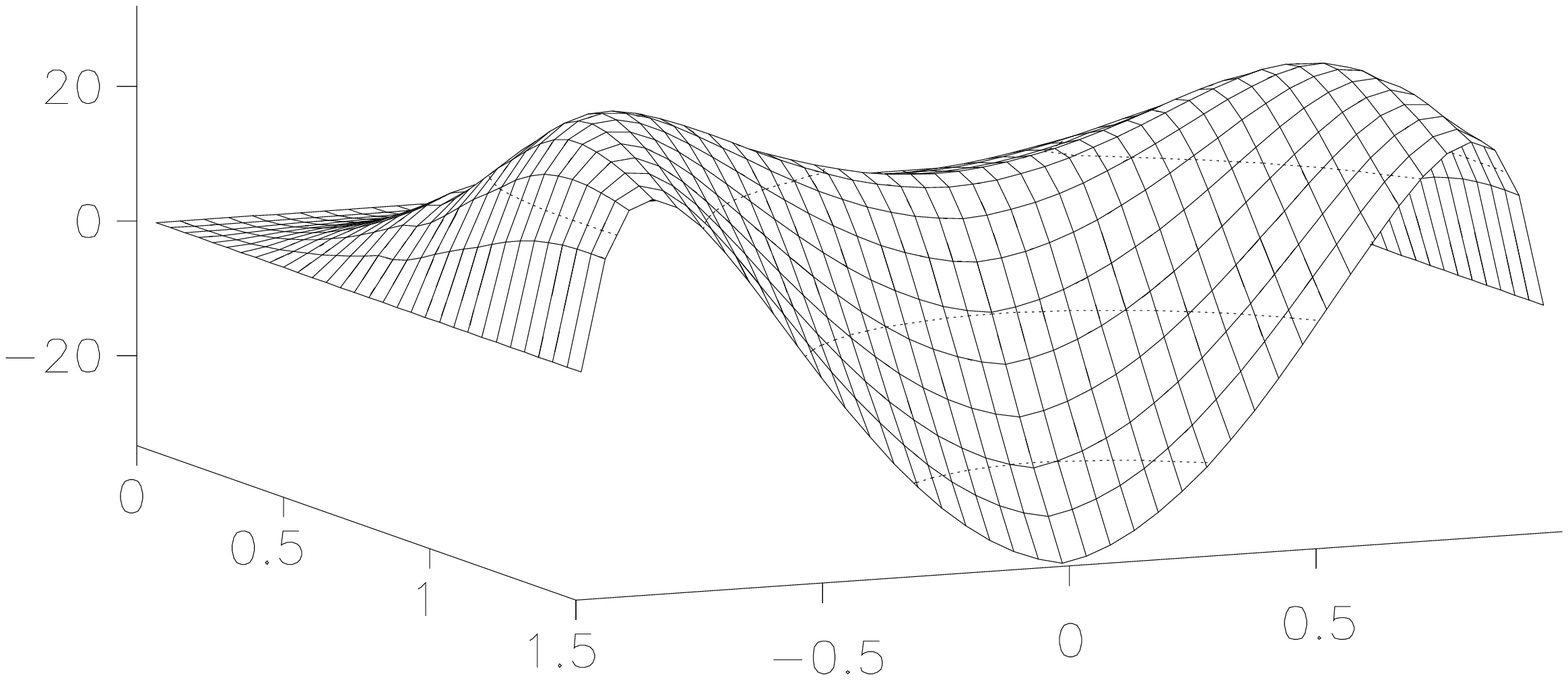}  
\end{center}
\caption{PES of $^{242}$Cm vs $(R-R_i)/(R_t - R_i)$ and $\eta =
(A_1-A_2)/(A_1+A_2)$. Y+EM (bottom), Shell + Pairing corrections (center), and
total deformation energy (top). The energies are 
expressed in MeV. \label{pescm}}
\end{figure}
\begin{figure}[hbt]
\includegraphics[width=6cm]{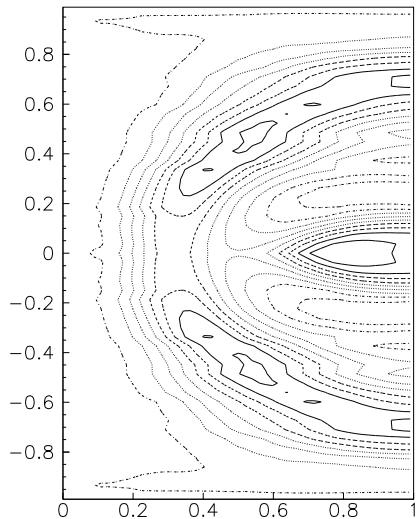}      
\caption{Contour plot of shell and pairing corrections for $^{242}$Cm  vs
$(R-R_i)/(R_t - R_i)$ and $\eta$.
\label{contcm}}
\end{figure}

The variation of the neutron and proton numbers of the two
fragments with the mass asymmetry at the touching point, $R=R_t$, for
$^{236}$Pu (figure~\ref{punzt}) shows the occurence of magic numbers
which generate the minima of the shell and pairing corrections producing the
valleys on PES. The very deep valley around $\eta =0.75$ ($\eta_A =0.763$,
$\eta_Z =0.745$) is produced by reaching simultaneously two magic
numbers $N_1=126$ and $Z_1=82$, $^{236}$Pu $\rightarrow \ ^{208}$Pb +
$^{28}$Mg. The cold fission valley ($^{236}$Pu $\rightarrow \ ^{132}$Sn +
$^{104}$Ru) at a small asymmetry ($\eta_A =0.119$, $\eta_Z =0.064$) is the
result of two magic numbers $Z_1=50$ and $N_1=82$. The intermediate valley
($\eta_A =0.288$, $\eta_Z =0.277$) is produced by the magicity of the light
fragment $N_2=50$, $^{236}$Pu $\rightarrow \ ^{152}$Nd +
$^{84}_{34}$Se$_{50}$.

\subsection{$^{242}$Cm}

There are four valleys on the PES of figure \ref{pescm} for $^{242}$Cm at
$\eta \simeq 0.7$; $0.4$; $0.3$, and $0.1$. They are also present 
on the corresponding contour plot (Fig.~\ref{contcm}).
At the top of Fig.~\ref{pescm} the deepest valley remains that of cold
fission due to a doubly magic heavy fragment $^{132}$Sn, 
\begin{figure}[hbt]
\includegraphics[width=6cm]{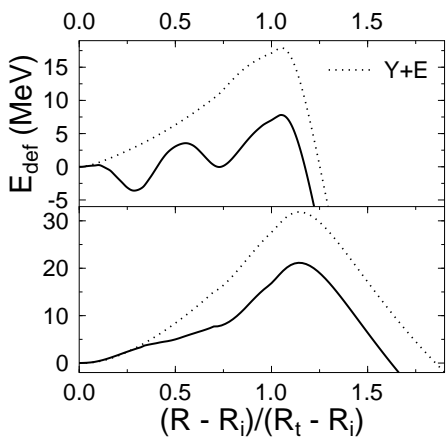}
\caption{A cut through the PES of $^{242}$Cm at symmetry $\eta =
0$ (top) and for $^{34}$Si radioactivity with $^{208}$Pb daughter (bottom). 
\label{barcm}}
\end{figure}
\begin{figure}[hbt]
\includegraphics[width=8cm]{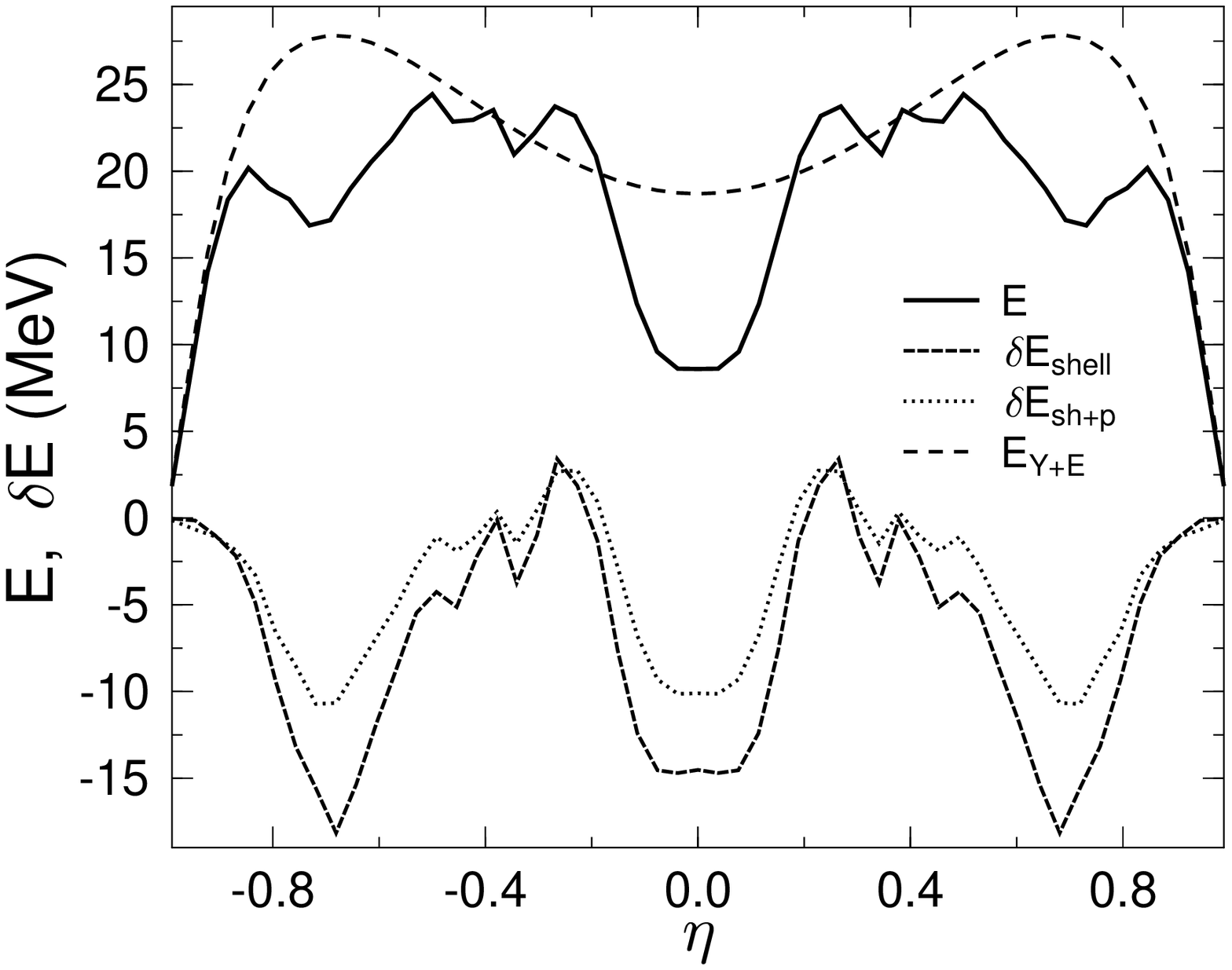}
\caption{Deformation energies at the touching point configurations ($R=R_t$)
of $^{242}$Cm vs the asymmetry $\eta $: $E_{Y+EM}$ and $E_{def}$ (top); shell
+ pairing corrections and only shell corrections (bottom). \label{cmtouc}}
\end{figure}
\begin{figure}[hbt]
\includegraphics[width=6cm]{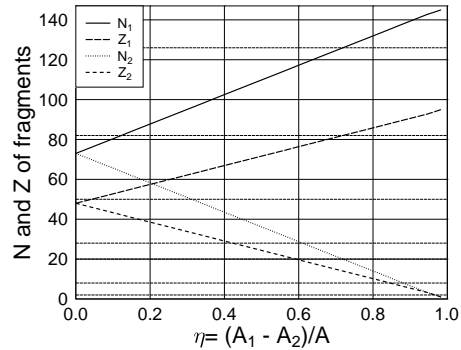}
\caption{Variation of the neutron and proton numbers of the two 
fragments with the mass asymmetry at the touching point, $R=R_t$ for
$^{242}$Cm. Every time a nucleon number reaches a magic value, the
corresponding shell correction has a local minimum. The very deep valley
around $\eta =0.719$ is produced by reaching simultaneously three magic
numbers $N_1=126$,  $N_2=20$, and $Z_1=82$. \label{cmnzt}}
\end{figure}
and the silicon-and-lead valley  the $^{208}$Pb + $^{34}$Si,
explaining the $^{34}$Si radioactivity of $^{242}$Cm.

Two plots obtained by cutting the PES at a given value of the asymmetry are
shown in Fig.~\ref{barcm}. 
In the upper part one can see the two humped barrier and at the bottom
the barrier for $^{34}$Si emission from $^{242}$Cm.
Having a
smaller height and width compared to the (dotted line) macroscopic Y+EM
barrier, it is very similar to the barrier used in ASAF which was lower and
narrower than the the Myers-Swiatecki's LDM barrier.

One can see how deep are the four valleys on the PES by plotting in
Fig.~\ref{cmtouc} a cut of the PES at the touching point configuration,
$R=R_t$. In the upper part of this figure the macroscopic energy,
$E_{Y+E}$, (the smooth dashed line) and the total deformation energy, $E$,
(heavy line) can be seen. 
In the lower part there are two other curves
representing the shell plus pairing correction (dotted curve), $\delta 
E_{sh+p}$, and only the shell corrections (dashed line), $\delta E_{shell}$. 

The figure ~\ref{cmnzt} shows the variation of the neutron and proton
numbers of the two
fragments with the mass asymmetry at the touching point, $R=R_t$, for
$^{242}$Cm. Like in the preceding subsections, the deepest valley on the
shell and pairing corrections is produced by the very strong shell effect of
the heavy fragment $^{208}$Pb ($^{242}$Cm
$\rightarrow \ ^{208}$Pb + $^{34}_{14}$Si$_{20}$) at $\eta_A =0.719$ and
$\eta_Z =0.708$; here there is also a contribution coming
from the neutron number of the light fragment, $N_2=20$.
The deepest valley on the total
deformation energy is the cold fission one at 
$\eta_A =0.091$, $\eta_Z =0.042$, where $^{242}$Cm $\rightarrow \
^{132}$Sn + $^{110}$Pd. The two intermediate valleys are produced mainly by
$N_2=50$ and $Z_2=28$. One has at $\eta_A =0.306$, $\eta_Z =0.292$ the decay
$^{242}$Cm $\rightarrow \ ^{158}$Sm + $^{84}_{34}$Se$_{50}$ and at 
$\eta_A =0.421$, $\eta_Z =0.417$ %the decay
$^{242}$Cm $\rightarrow \ ^{172}$Er + $^{70}_{28}$Ni$_{42}$.

In conclusion the strong shell effect associated to the doubly magic
character of the daughter $^{208}$Pb, observed in the systematic analysis of
experimental results \cite{p240pr02}, comes from a valley present on the
potential energy surfaces of cluster emitters at a relatively high value of
the asymmetry parameter $\eta \simeq 0.7-0.8$. Despite its high depth on the
microscopic corrections PES, on the total deformation energy it appears
shallower since it is added to the Businaro-Gallone mountain of the Y+EM
macroscopic energy.
The potential barrier shape
of heavy ion radioactivity obtained from the first time by using the
macroscopic-microscopic method provides further support for the particular
choice of the barrier within the analytical superasymmetric model.
The depth of the cold fission valley produced by the doubly magic heavy
fragment $^{132}$Sn, which is small for $^{222}$Ra, increases with the mass
number of the parent nucleus being comparable to the lead valley on the
microscopic corrections PES of $^{242}$Cm. Some other intermediate shallower 
valleys are produced by the magicity of the light fragment: $Z_2=28$ for
$^{222}$Ra parent; $N_2=50$ for $^{232}$U parent; $N_2=50$ for $^{236}$Pu;
$N_2=50$ and $Z_2=28$ for $^{242}$Cm.

\begin{acknowledgments}
This work was partly supported by Deutsche Forschungsgemeinschaft, Bonn,
and by Ministry of Education and Research, Bucharest. We acknowledge also
the support by Prof. S. Hofmann Gesellschaft f\"{u}r Schwerionenforschung
(GSI), Darmstadt.
\end{acknowledgments}


\begin{thebibliography}{99}

\bibitem{p240pr02}
D.N. Poenaru,  Y. Nagame, R.A. Gherghescu, W. Greiner,
{\it Phys.  Rev.}, {\bf C65} (2002) 054308; Erratum {\bf C66} 049902. 

\bibitem{p195b96b}
D.N. Poenaru, W. Greiner,  in {\em Nuclear Decay Modes}, (IOP
Publishing, Bristol, 1996), Chap.~6, pp. 275.

\bibitem{p160adnd91}
D. N. Poenaru, D. Schnabel, W. Greiner, D. Mazilu, R. Gherghescu, {\it 
Atomic Data Nucl. Data Tables}, {\bf 48}  (1991) 231.

\bibitem{rad03prc}
R. A. Gherghescu, {\it Phys. Rev.} {\bf 67} (2003) 014309.

\bibitem{str67np}
V. M. Strutinsky, {\it Nucl. Phys.} {\bf A 95} (1967) 420.

\bibitem{sch69zp}
W. Scheid, W. Greiner, {\it Z. Phys.} {\bf A 226} (1969) 364.

\bibitem{kra79pr}
H. J. Krappe, J. R. Nix, A. J. Sierk, {\it Phys. Rev.} {\bf C 20} (1979)
992.

\bibitem{p80cpc80}
D. N. Poenaru, M. Iva\c{s}cu, D. Mazilu, {\it Comput. Phys. Comm.} {\bf 19}
(1980) 205.

\bibitem{mol95adndt}
P. M\" oller, J. R. Nix, W. D. Myers, and W. J. Swiatecki,
{\it Atomic Data Nucl. Data Tables} {\bf 59} (1995) 185. 

\bibitem{bol72pr}
M. Bolsterli, E. Fiset, J. R. Nix, J. Norton, {\em Phys. Rev} {\bf C 5}
(1972) 1050.

\bibitem{mye66np}
W. D. Myers, W. J. Swiatecki, {\em Nucl. Phys.} {\bf A 81} (1966) 1.

\end{thebibliography}
\end{document}